# Survey and Test Environment for ITER EPP#12 In-PP Electrical Components

X. Sun, *Member, IEEE*, F. Wang, *Member, IEEE*, Q. Hu, C. Xu and M. Nie

*Abstract*—The purpose of Equatorial Port Plug 12 (EPP#12) for International thermonuclear experimental reactor (ITER) is to provide a common platform and interface, support or constrainer for five diagnostic plant systems and one glow discharging cleaning system (GDC). As EPP#12 integrator, a team from Institute of plasma physics Chinese of Sciences (CASIPP) performs the design work. The Instrument and Control(I&C) is an important part of system design. The main I&C functions will be implemented include temperature measurements of the port structures, electrical heater with temperature control during baking of windows and providing spare input measurement channel. The integrator should provide the embedded temperature sensors, associated cabling, electrical connectors and electrical feedthrough. Most electrical components will be deployed in port plug structure which is a harsh environment for electrical components. In this paper, we present the survey and research of electrical components for ITER EPP#12. And the design and implement of a test environment for electrical components which is based-on ITER CODAC is also described.

## I. INTRODUCTION

THE ITER[1] facility, being built as an international collaboration in southern France, will be the world's largest magnetic confinement plasma physics experiment and explore a plasma parameter envelope currently not available in tokamaks.

Diagnostic plant systems which provide the means to observe control and sustain the plasma performance are critical part of the operation of ITER [2]. There are 8 equatorial port plugs and 10 upper port plugs in vacuum vessel to provide support structure and common platform for diagnostic systems and to provide common interfaces to other plant systems, such as cooling water, remote handling, liquid & gas distribution system and so on. As one of the two port plug s for ITER "Phase 1", EPP#12 is designed to contain five diagnostic systems and one GDC system. It will be a challenging task to integrate the tenant systems. The Instrument and Control (I&C) system is an important part of EPP#12 design work. The main purpose of EPP#12 I&C is to measure and monitor the environment parameters in the port plug and to ensure the operating safety not only the port plug but also the tenant systems. In port plug, most I&C instruments and components will be deployed in a high radiation, vacuum and temperature environment. To ensure the accuracy and reliability of measurement, the selection of electrical components is the key task for EPP#12 I&C.

The purpose of EPP#12 consists of three assemblies, port plug structure (PPS), interspace support structure (ISS) and port cell support structure (PCSS). Closure Plate is the vacuum boundary component between PPS and ISS.

The purpose of it is to integrate and provide stable platform for five diagnostics and one GDC system, the equatorial port visible/IR wide angle viewing system (Vis/IR), H-alpha and visible spectroscopy system, radial x-ray camera, hart x-ray monitor system and collective Thomson scattering diagnostic (CTS) and glow discharge cleaning system. The tenant systems are located in five regions with different vacuum, radiation, temperature and have to share space and support services with each other.

The port plug structure integrated with three Diagnostic Shield Modules (DSMs) is the most important part of EPP#12. At current stage, the I&C work for EPP#12 focuses on the selection and R&D of electrical components in port plug structure (In-PP). In PPS, the integrator should design and provide temperature sensors (thermocouples), associating cabling, electrical connectors attached to DSMs and electrical feedthrough, cables, connectors attached to closure plate.

All In-PP electrical components have to work in a harsh environment. The main performance requirements are listed as follows: Work temperature: 70°C (operation), 240°C (baking); Vacuum Quality Classification [3]: VQC1A, the highest level; Radiation: neutron flux: 1011n/cm2/s; gamma dose: 100 mGy; Remote Handling Compatibility

The thermostability, radioactivity resistant and gas-tightness are key technical issues for the selection and R&D of EPP#12 electrical components. The survey and R&D was triggered by the requirements to select temperature sensors, connectors and associated cabling for EPP#12 and will focus on material , structure，isolation and insulation.

The measurement of environment temperature is a main function of EPP#12 I&C. The thermocouples are mounted at selected location on the port structure s for monitoring purpose. The main In-PP electrical components include thermocouple, extension cables, cable routing, electrical connector and feedthrough. The sensors, cables and other electrical components must be in full     compliance with the rules and guidelines proposed by ITER organization (ITER IO

The thermocouples will be placed in a high radiation environment in PPS.  According to the Chinese technical specification for nuclear grade quality sheathed thermocouple (EJ 660-92) [5], thermocouple type, sheath material, insulated hot junction type and insulation material type should be taken into consideration.

For better thermoelectric stability and reducing foreign matter, for example manganese, the type N(NiCrSi+-NiSi-) thermocouples was selected for ITER EPP#12. Type N thermocouples are suitable for use between -270 °C and 1300 °C. Sensitivity is about 39 μV/°C at 900 °C, slightly lower compared to type K.

The vacuum level in PP is VQC1A. The accepted materials in VQC1A include 316L, 316LN and 316L(N)-IG. The Sheathed material for the thermocouples should be 316L or 316LN stainless steel (SS). 316L SS is more common. It is low carbon content stainless steel, easy to weld and corrosion and heat resistant steel. It has good resistance against a variety of aggressive media and small sensitivity against intra-crystalline corrosion because of the low carbon content. The maximum temperature for continuous utilization is 800°C. 316L SS sheathed thermocouples can be used for nuclear energy research and development.

The standard insulated hot junction type is selected. The hot junction is insulated from the sheath. Minerals such as magnesium oxide (MgO) and aluminum oxide ($Al_2O_3$) are the most suitable insulating material for mineral insulated thermocouple. The pureness of insulation material should be more than 99.5%.

Thermocouple is connected to the measurement device by means of an extension cable. The key feature should be mindful for extension cable selection are listed as below: Working temperature range; Chemical resistance of sheath material ; Abrasion and vibration resistance; Installation requirements; Vacuum requirements

The working temperature will be 70°C during operation and will be 240°C during baking. To avoid measurement errors, the extension cables should be the same as for the thermocouple. The extension cables should be radiation and fire resistance. High vacuum compatible should be taken into consideration. Only Mineral insulated (MI) cable and Optical fibre metal coating cable can be used in VQC1A environment. 316SS, Pure copper and magnesium Oxyde are available material for MI cable sheath, conductor and insulator material.

We got some sample of thermocouple and extension cables from Okazaki company [6] for testing.

Electrical Connector

There are three types of connectors should be defined in PP. The first kind is for High Power/High Frequency (HP/HF). The High Frequency means more than 100MHz and impedance is 50ohm. Voltage is less than 1kV. This type of cable is used as the power supply cable for CTS in DSM3. The second is designed for Middle Power/Low Frequency (MP/LF). It will be used to connect Shielded twisted pair cable for Light sources of Vis/IR. The third one is for Low Power/Low Frequency (LP/LF), which is used to connect shielded twisted pair and thermocouple. All the connectors should have remote handling compatible. There are no commercial connector products can satisfy the requirements of In-PP connector. The R&D work of multi-types connector is essential. A spiral deployment structure will be used.

At the cold end, the thermocouple must be fitted with a suitable connector to connect to the extension cable. There are more than 100 thermocouples will be placed in 3 DSMs for 6 tenant systems and port plug. The size, material and installation (where and how) and remote handling will be main factors for thermocouple connector design. We compared some thermo connectors from different manufacturers. LEMO [6] S series provides push-pull self-locking thermo connector which support remote handling. Though LEMO didn't provide the connector with N type thermo-contact, a special golden version can be mounted on the most used thermocouples as mentioned in reference [6]. The other thermo-connector product comes from Lasker company. There are 7 connect channel for N type thermocouple in one connector. But the size is big and doesn't have remote handling compatible. At next stage, the integrator will focus on the mechanical design of Thermo-connector for N type thermocouples in PPS.

To meet shielding and dry weight requirement of total EPP assembly, a new modular design which using $B_4C$ as shielding material will replace the original water shielding model. The $B_4C$ blocks will be installed on the vertical plate structure which will significantly affect the cable routing deployment.

Most of the cables used in PPS are mineral insulated cables. The bending radius of MI cable is much larger than common cables. MI cables doesn't have good flexibility. The rectangle shape cable routing proposal is unfit for mineral cable bundle in PPS. The straight channel cable routing proposal which slotting in top and bottom plate will be accepted.

Electrical feedthrough will provide an external servers interface at the closure plate. There will be 6 electrical feedthroughs in closure plate. To save space, integrator will not design dedicated electrical feedthrough and will share with tenant systems. As the vacuum boundary component, the electrical feedthrough should be in conformance with RCC-MR and IEE 317 standards and have the capability of being baked to 200°C and remote handling.

The R&D work of electrical feedthrough is still ongoing. It will take in a year to complete.

To test the electrical components performance, for example fatigue test, a test environment has been setup. The main function of the test environment is to test the performance of electrical components under different condition (temperature). The test environment is consisting of four dedicate components, a programmable electrical heater, a PLC for analogue input data acquisition, an Industrial computer with ITER CODAC system for data storage and human interface and a network switch for data transfer.

The thermocouples and extension cable samples work well in normal temperature and baking temperature environment. In Fig.6., the human interface for test application is described.

## DISCLAIMER

The opinions and technical proposal presented here in do not necessarily reflect those of the ITER organization.

## REFERENCES

[1] http://www.iter.org